\documentclass[journal]{IEEEtran}
\usepackage{cite}
\usepackage{amsmath,amssymb,amsfonts,amsthm}
\usepackage{algorithmic}
\usepackage{graphicx}
\usepackage{textcomp}
\usepackage{caption}
\usepackage{subfigure}
\usepackage{booktabs}
\usepackage[colorlinks=true, allcolors=blue]{hyperref}
\usepackage[linesnumbered,ruled,vlined]{algorithm2e}
\usepackage{wrapfig}
\usepackage{cite}
\usepackage{xr}
\usepackage{hyperref}
\usepackage{bm}
\usepackage[normalem]{ulem}
\DeclareMathOperator*{\argmax}{argmax}
\DeclareMathOperator*{\argmin}{argmin}
\makeatletter
\newcommand*{\addFileDependency}[1]{
  \typeout{(#1)}
  \@addtofilelist{#1}
  \IfFileExists{#1}{}{\typeout{No file #1.}}
}
\makeatother
\newcommand*{\myexternaldocument}[1]{
    \externaldocument{#1}
    \addFileDependency{#1.tex}
    \addFileDependency{#1.aux}
}

\myexternaldocument{Tdenoisesupplemental}

\usepackage{adjustbox}
\usepackage{setspace}
\usepackage{hyperref}
\usepackage{multirow}
\usepackage[switch,columnwise]{lineno}
\usepackage{chapterbib}

\newcommand{\etal}{\textit{et al}.}

\SetKwInput{KwInput}{Input}                
\SetKwInput{KwOutput}{Output}              

\begin{document}

\title{A Hybrid Approach for Approximating the Ideal Observer for Joint Signal Detection and Estimation Tasks by Use of Supervised Learning and Markov-Chain Monte Carlo Methods}

\author{Kaiyan Li, \IEEEmembership{Student Member, IEEE}, Weimin Zhou, Hua Li, and Mark A. Anastasio \IEEEmembership{Senior Member, IEEE}
\thanks{
Kaiyan Li is with the Department of Bioengineering, University
	of Illinois at Urbana-Champaign, Urbana, IL, 61801 USA (Email:
	kaiyanl2@illinois.edu).
Weimin Zhou was with the Department of Electrical and Systems
	Engineering, Washington University in St. Louis, St. Louis, MO, 63130
	USA. He is now with the Department of Psychological and Brain Sciences, University of California, Santa Barbara, Santa Barbara, CA 93106 USA (Email: weiminzhou@ucsb.edu).
Hua Li is with the Department of Radiation Oncology, Washington University School of Medicine, Saint Louis, MO,
She was with the Department of Bioengineering, University of Illinois
	at Urbana-Champaign, Urbana, IL 61801 USA. (Email:li.hua@wustl.edu \& huali19@illinois.edu).
Mark A. Anastasio is with the Department of Bioengineering, University
	of Illinois at Urbana-Champaign, Urbana, IL, 61801 USA (Email:
	maa@illinois.edu).
This work was supported in part by NIH awards R01EB020604, R01EB023045, R01NS102213, R01CA233873, and R21CA223799. (Corresponding authors: Mark A. Anastasio \& Hua Li).
}
\vspace{-0.6cm}
}
\maketitle

\begin{abstract}
The ideal observer (IO) sets an upper performance limit among all observers and has been advocated for assessing and optimizing imaging systems.
For general joint detection and estimation (detection-estimation) tasks, estimation ROC (EROC)  analysis has been
established for evaluating the performance of observers. 
However, in general, it is difficult to accurately approximate the IO that maximizes the
area under the EROC curve.
In this study, a hybrid method that employs machine learning is proposed to accomplish this.
Specifically, a hybrid approach is developed that combines a multi-task convolutional
 neural network and a Markov-Chain Monte Carlo (MCMC) method in order to approximate the IO for
detection-estimation tasks.
Unlike traditional MCMC methods, the hybrid method is not limited to use of specific utility functions.
In addition, a purely supervised learning-based sub-ideal observer is proposed.
Computer-simulation studies are conducted to validate the proposed method, which include
signal-known-statistically/background-known-exactly and signal-known-statistically/background-known-statistically tasks.
The EROC curves produced by the proposed method are compared to those produced by the MCMC approach or analytical computation when feasible.
The proposed method provides a new approach for approximating the IO and may advance the application of 
EROC analysis for optimizing imaging systems.
\end{abstract}


\section{Introduction}
\label{sec:introduction}  
Objective, or task-based, measures of image quality (IQ) are  advocated  for use in the assessment and optimization of medical imaging systems \cite{metz1995toward,barrett1993model,barrett2013foundations,he2013model,abbey2000modeling}.
Unlike traditional physical measures of IQ, task-based measures of IQ
quantify the ability of an observer to perform a specific task such as detection or estimation of a signal. 
To compute such measures of IQ when developing and refining new imaging technologies,
numerical observers (NOs) have been widely employed \cite{he2013model,barrett1993model}. This enables the exploration of large parameter spaces when optimizing system performance. 
Such NOs can be designed to estimate an upper bound on the possible performance of any observer for a given task and collection of image data. Such models are referred to as \emph{ideal observers} (IOs). Alternatively, anthropomorphic NOs can be designed  to mimic the performance of a human observer, which is generally sub-optimal. The focus of this work will be on the computation of IOs.


The Bayesian IO is a numerical observer (NO) that achieves the performance of the optimal decision maker acting on given measured data \cite{barrett2013foundations,clarkson2007estimation,li2021supervised}. 
The IO performance is a task-based image quality measure that depends on the data and the task (e.g., lesion detection) and not the capabilities of a human observer, the quality of feature extraction, or a particular classification scheme. As such, it is a metric of fundamental importance in the objective assessment of imaging hardware and data-acquisition designs \cite{metz1995toward}. 
Knowledge of the IO performance is also important because it can reveal when task-related information is readily extracted by a human observer or another sub-optimal NO. 
This can permit the identification of opportunities for improved image processing or other methodological changes that lead to improved task-performance.

Much of the literature on IO approximation has focused on binary signal detection tasks.
The IO test statistic in this case is generally a non-linear function of
the image data and, except in some special cases, cannot be determined analytically. 
Because of this, sampling-based methods that employ Markov-chain Monte Carlo (MCMC) techniques and supervised learning have been developed to approximate the IO test statistic for medical image applications \cite{kupinski2003ideal,zhou2019approximating,zhou2020approximating}.

Detection-estimation tasks that involve the detection of a signal and the subsequent estimation of a set of parameters for the signal present decisions are relevant to many medical imaging applications \cite{clarkson2007estimation,wunderlich2014nonparametric,whitaker2008estimating,lee2013assessment,tseng2017assessing,zhang2013channelized}.
The estimation receiver operating characteristic (EROC) curve \cite{clarkson2007estimation} can be employed to assess the performance of an observer on detection-estimation tasks and the area under the EROC curve (AEROC) can be utilized as a figure-of-merit (FOM). Similar to the case of binary detection tasks, the IO decision strategy for detection-estimation tasks is analytically intractable except in special cases \cite{wunderlich2016optimal}.
However, unlike the  of case binary detection tasks for which MCMC methods or supervised learning methods can be employed to establish NOs, \cite{kupinski2001ideal,zhou2019approximating}, there is a lack of available NOs for approximating the IO for detection-estimation tasks.

To address this need, in this study a hybrid method is developed for approximating the IO 
for a wide class of detection-estimation tasks. The proposed method
combines deep learning (DL) and a Markov-Chain Monte Carlo (MCMC) method in order to implement the known IO decision strategy \cite{clarkson2007estimation}.
Unlike traditional MCMC methods for approximating the IO, the hybrid method is not limited to use of specific utility functions.
In addition, a purely supervised learning-based sub-ideal observer is designed, which avoids the need to involve the MCMC method.
Computer-simulation studies are conducted to validate the performance of the proposed methods, 
which address signal-known-statistically/background-known-exactly (SKS/BKE) and signal-known-statistically/background-known-statistically (SKS/BKS) tasks.
The proposed methods provide a new capability for approximating the IO for detection-estimation tasks and may advance the application of 
EROC analysis for optimizing imaging systems.


The remainder of the paper is organized as follows. 
Section~\ref{sec:background} provides salient information regarding  signal detection-estimation theory. The proposed hybrid IO approximation method and the supervised learning-based sub-ideal observer are described in Section~\ref{sec:method}.
The numerical studies and the results are provided in Sections~\ref{sec:numerical_studies} and~\ref{sec:results}, respectively.
Finally, the article concludes with a discussion in Section~\ref{sec:discussion}.

\section{Background}
\label{sec:background}
A linear digital imaging system can be described as a continuous-to-discrete (C-D) mapping process \cite{barrett2013foundations}: 
\begin{equation}
\mathbf{g}=\mathcal{H}{f(\mathbf{r})+\mathbf{n}},   
\label{eq:imaging_system}
\end{equation}

\noindent where $\mathbf{g}\in\mathbb{R}^{N\times 1}$ is the measured image vector, 
$f(\mathbf{r})$ denotes the object function that is dependent on the coordinate $\mathbf{r}\in\mathbb{R}^{k\times 1}$, $k \ge 2$,
$\mathcal{H}$ denotes a linear imaging operator that maps $\mathbb{L}_{2}(\mathbb{R}^{k})$ 
to $\mathbb{R}^{N\times 1}$, 
and $\mathbf{n}\in\mathbb{R}^{N\times 1}$ denotes the measurement noise. 
The imaging process described in Eqn. (\ref{eq:imaging_system}) can be expressed as \cite{barrett2013foundations}: 
\begin{equation}
[\mathbf{g}]_m\equiv \int_V  d\mathbf{r}\,f(\mathbf{r})\,h_m(\mathbf{r})+[\mathbf{n}]_m,
\label{eq:imaging_system_2}
\end{equation}
where $[\mathbf{g}]_m$ and $[\mathbf{n}]_m$ denote the $m^{th}$ component of $\mathbf{g}$ and $\mathbf{n}$, respectively, $V$ denotes the support of $f(\mathbf{r})$, and $h_m(\mathbf{r})$ is the point response function (PRF) of the imaging system \cite{barrett2013foundations}. 

\subsection{General detection-estimation tasks}
\label{ssec:eroc_general}
In a binary detection-estimation task, each image is either signal-absent or contains a signal that is specified by a parameter vector $\bm{\theta}$. The imaging processes under these hypotheses can be expressed as:
\begin{subequations}
	\label{eq:hypo}
	\begin{equation}
	H_0:\mathbf{g}=\mathcal{H}\mathbf{f_b}+\mathbf{n}=\mathbf{b+n}, 
	\end{equation}
	\begin{equation}
	H_1:\mathbf{g}=\mathcal{H}\mathbf{(f_b+f_{s(\bm{\theta})})}+\mathbf{n}=\mathbf{b+s(\bm{\theta})+n},   
	\end{equation}   
\end{subequations}
where $\mathbf{f_{s(\bm{\theta})}}$ and $\mathbf{f_b}$ denote the signal and background, respectively, and $\mathbf{s}\equiv\mathcal{H}\mathbf{f_{s(\bm{\theta})}}$ and $\mathbf{b}\equiv\mathcal{H}\mathbf{f_b}$ denote the signal and background images.
To perform this task, a deterministic observer first
computes a test statistic $T(\mathbf{g})$ that maps the measured image $\mathbf{g}$ to a real-valued scalar variable. The value of $T(\mathbf{g})$ is then compared to a
predetermined threshold $\tau$ to determine if $\mathbf{g}$ satisfies $H_0$ or $H_1$.
Subsequently, an estimate $\bm{\hat{\theta}}(\mathbf{g})$ is given if the observer decides that the signal is present \cite{clarkson2007estimation,wunderlich2016optimal}. 

An EROC curve is generated by plotting the expected utility of $\bm{\hat{\theta}}(\mathbf{g})$ for the true positive (TP) decisions versus the false-positive (FP) fraction as $\tau$ is varied. 
The expected utility of $\bm{\hat{\theta}}(\mathbf{g})$ for a true positive (TP) decision (referred to as the expected utility) is defined as \cite{clarkson2007estimation}
\begin{equation}
U_{TP}(\tau)=E[u(\hat{\bm{\theta}}(\mathbf{g}),\bm{\theta})step[T(\mathbf{g})-\tau]|H_1],
\label{eq:U_TP}
\end{equation}
where $U_{TP}(\tau)$ is the expected utility, $u(\hat{\bm{\theta}}(\mathbf{g}),\bm{\theta})$ is a utility function for the parameter estimate, $E[\cdot]$ is the mathematical expectation operator, and $step[\cdot]$ is the Heaviside step function. In general, the utility function should  be designed to return a high value when $\hat{\bm{\theta}}(\mathbf{g})$ is close to $\bm{\theta}$ and a low value otherwise.
Each point on the EROC curve gives the expected utility of the estimated parameter for the TP decisions at a given false-positive fraction. Because the utility function may be negative, unlike the traditional ROC curve, the EROC curve  is not always an increasing as a function of FPF.

\subsection{Ideal observer for detection-estimation tasks}
\label{IO_background}
The IO test statistic $T_I(\mathbf{g})$ and estimator $\hat{\bm{\theta}}_I(\mathbf{g})$ can be computed as \cite{clarkson2007estimation}: 
\begin{equation}
\label{eq:IO_test_estimate_t_stat}
T_I(\mathbf{g})=\max_{\hat{\bm{\theta}}}\left\{\int  p(\bm{\theta})\Lambda(\mathbf{g}|\bm{\theta})u(\bm{\theta},\hat{\bm{\theta}})d\bm{\theta}\right\},
\end{equation}
and
\begin{equation}
\label{eq:IO_test_estimate_est}
\hat{\bm{\theta}}_I(\mathbf{g})=\argmax_{\hat{\bm{\theta}}} \left\{\int p(\bm{\theta})\Lambda(\mathbf{g}|\bm{\theta})u(\bm{\theta},\hat{\bm{\theta}})d\bm{\theta}\right\},
\end{equation}
where the quantity $\Lambda(\mathbf{g}|\bm{\theta})$ is the $\bm{\theta}-$conditional likelihood ratio:
\begin{equation}
\Lambda(\mathbf{g}|\bm{\theta})=\frac{p(\textbf{g}|\bm{\theta},H_1)}{p(\textbf{g}|H_0)}.
\end{equation}
Equation (\ref{eq:IO_test_estimate_t_stat}) implies that the IO test statistic can be written as \cite{clarkson2007estimation}:
\begin{equation}
\label{eq:IO_test_2}
T_I(\mathbf{g})=\int  p(\bm{\theta})\Lambda(\mathbf{g}|\bm{\theta})u(\bm{\theta},\hat{\bm{\theta}}_I(\mathbf{g}))d\bm{\theta},
\end{equation}
where $\hat{\bm{\theta}}_I(\mathbf{g})$ is the ideal estimator defined in Eqn. (\ref{eq:IO_test_estimate_est}).

\subsection{Scanning linear observer}
\label{ssec:slo}
The scanning linear observer (SLO) \cite{whitaker2008estimating} is a sub-optimal linear observer that can be employed with detection-estimation tasks when the IO is intractable \cite{tseng2017assessing,lee2013assessment}. The SLO is designed to approximate the mode of the posterior density and perform pseudo maximum a posteriori (MAP) estimation \cite{whitaker2008estimating,tseng2017assessing} as:
\begin{equation}
\label{eq:SLO_estimator}
\begin{aligned}
\hat{\bm{\theta}}_{SLO}(\mathbf{g})=&\argmax_{\hat{\bm{\theta}}}\{ \mathbf{\bar{g}(\hat{\bm{\theta}})}^T\mathbf{\bar{K}_g}^{-1}\mathbf{g}-\frac{1}{2}\mathbf{\bar{g}(\hat{\bm{\theta}})}^T\mathbf{\bar{K}_g}^{-1}\mathbf{\bar{g}(\hat{\bm{\theta}})} \\
&+\ln p(\hat{\bm{\theta}}) \},
\end{aligned}
\end{equation}
where $\mathbf{\bar{g}}(\hat{\bm{\theta}})$ is the mean image averaged over the parameters and $\mathbf{\bar{K}_g}^{-1}$ is the inverse covariance matrix for $(\mathbf{g}-\mathbf{\bar{g}(\hat{\bm{\theta}})})$ with approximation of slowly varying parameters. The corresponding SLO test statistic is given by
\begin{equation}
\label{eq:SLO_test_stat}
\begin{aligned}
T_{SLO}(\mathbf{g})=\,& \mathbf{\bar{g}}(\hat{\bm{\theta}}_{SLO}(\mathbf{g}))^T\mathbf{\bar{K}_g}^{-1}\mathbf{g}\\
&-\frac{1}{2}\mathbf{\bar{g}}(\hat{\bm{\theta}}_{SLO}(\mathbf{g}))^T\mathbf{\bar{K}_g}^{-1}\mathbf{\bar{g}}(\hat{\bm{\theta}}_{SLO}(\mathbf{g}))\\
&+\ln p(\hat{\bm{\theta}}_{SLO}(\mathbf{g})).
\end{aligned}
\end{equation}

\section{Approximating the IO for signal detection-estimation tasks}
\label{sec:method}
\subsection{Approximating the IO using a hybrid supervised learning-MCMC method}
The supervised learning-based methods employed in previous studies \cite{kupinski2001ideal,zhou2019approximating,zhou2020approximating} cannot be utilized directly to approximate the IO test statistic for detection-estimation tasks.
One reason is that the test statistic $T_I(\mathbf{g})$ in Eqn. (\ref{eq:IO_test_2}) involves an integral that depends on the ideal estimator $\hat{\bm{\theta}}_I(\mathbf{g})$.  
To circumvent this, a novel hybrid supervised learning-Markov-Chain Monte Carlo (MCMC) method is described below for IO approximation.

Considering that the signal parameter vector $\bm{\theta}$ is estimated only when the signal is determined to be present, $p(\bm{\theta})=p(\bm{\theta} |H_1)$ holds true \cite{clarkson2007estimation,wunderlich2016optimal}. Thus, the test statistic $T_I(\mathbf{g})$ can be expressed as
\begin{equation}
T_I(\mathbf{g})= \int p(\bm{\theta} |H_1)\frac{p(\mathbf{g}|\bm{\theta},H_1)}{p(\mathbf{g}|H_0)}u(\hat{\bm{\theta}}_I(\mathbf{g}),\bm{\theta})d\bm{\theta}.
\end{equation}
Using Bayes's rule, $T_I(\mathbf{g})$ can then be decomposed as \cite{li2021supervised}
\begin{equation}
\begin{aligned}
T_I(\mathbf{g})
&= \int \frac{pr(\mathbf{g}|H_1)}{pr(\mathbf{g}|H_0)}pr(\bm{\theta}|\mathbf{g},H_1)u(\hat{\bm{\theta}}_I(\mathbf{g}),\bm{\theta})d\bm{\theta} \\
&\equiv \Lambda(\mathbf{g})U(\mathbf{g}),
\label{eq:t_decompose}
\end{aligned}
\end{equation}
where $ \Lambda(\mathbf{g})\equiv \frac{p(\mathbf{g}|H_1)}{p(\mathbf{g}|H_0)}$ is the likelihood ratio and 
\begin{equation}
\label{eq:U_g}
U(\mathbf{g})\equiv\int pr(\bm{\theta}|\mathbf{g},H_1)u(\hat{\bm{\theta}}_I(\mathbf{g}),\bm{\theta})d\bm{\theta}
\end{equation}
is the utility weighted posterior mean. Equation (\ref{eq:t_decompose}) is central to the methodology described below.

\begin{figure}[ht]
	\centering
		\includegraphics[width=0.5\textwidth]{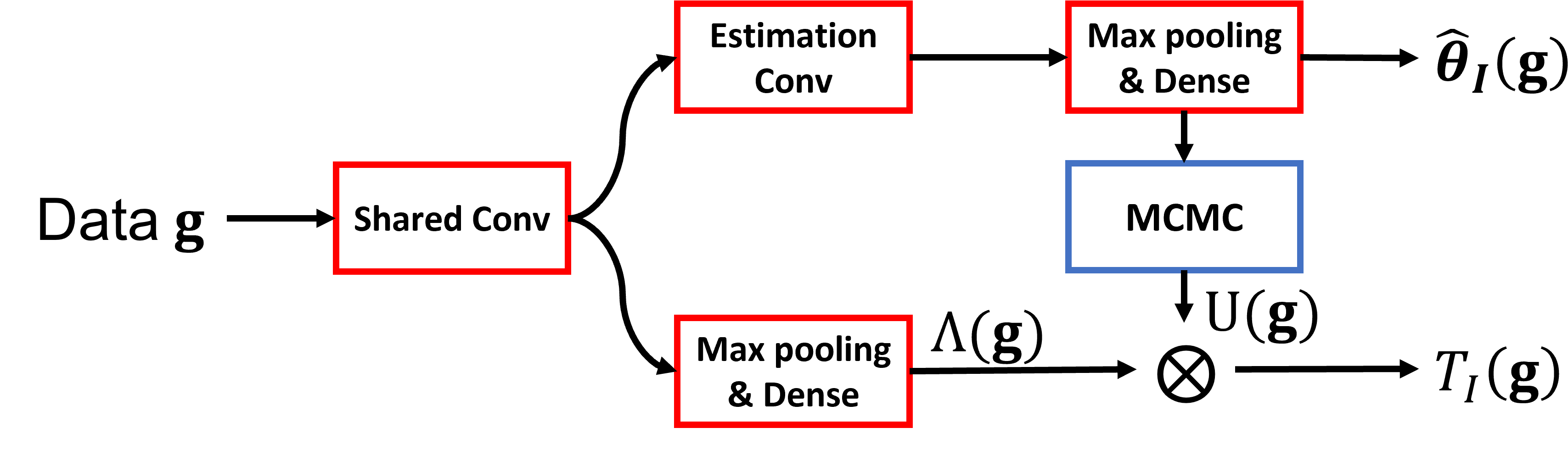}
	\caption{A schematic of the hybrid method that combines a multi-task CNN and the MCMC method. The Shared Conv and Estimation Conv are two blocks that comprise several convolutional layers. The convolutional layers in the Shared Conv 
	are shared by the detection and estimation sub-networks  (i.e., the bottom and top network branches). The layers in the Estimation Conv are employed by the estimation sub-network only.}
	\vspace{-0.1in}
    \label{fig:IO_architecture}
\end{figure}

The proposed hybrid supervised learning-MCMC strategy is summarized in Fig. \ref{fig:IO_architecture}. First, a multi-task CNN is employed to approximate the likelihood ratio $\Lambda(\mathbf{g})$ and the ideal estimate $\hat{\bm{\theta}}_I(\mathbf{g})$. Second, $U(\mathbf{g})$ is approximated by use of a MCMC technique. Finally, the IO test statistic is obtained by multiplying $\Lambda(\mathbf{g})$ and $U(\mathbf{g})$ according to Eqn. (\ref{eq:t_decompose}). The corresponding details are described below.

\subsubsection{Approximating the likelihood ratio and ideal estimate using multi-task CNNs}
\label{sssec:multi-task}
The likelihood ratio $\Lambda(\mathbf{g})$ and the ideal estimate $\hat{\bm{\theta}}_I(\mathbf{g})$ can be approximated by use of CNNs.
Specifically, as shown in Fig. \ref{fig:IO_architecture}, a multi-task CNN architecture is employed
in which several convolutional layers in the Shared Conv block are shared by the detection and estimation sub-networks (i.e., the bottom and top network branches in Fig. \ref{fig:IO_architecture}). Several additional convolutional layers in the Estimation Conv block are employed by the estimation sub-network only.
For use in training the detection network, the loss function is defined as the binary cross entropy loss function to approximate the posterior probability $p(H_1|\mathbf{g})$, which is a monotonic transformation of $\Lambda(\mathbf{g})$ \cite{zhou2019approximating}. 
For use in training the estimation sub-networks, the loss function is defined as the negative of the utility function \cite{clarkson2007estimation,clarkson2019risk}. The derivation of the loss function for estimation tasks is provided in  Appendix \ref{ap:est_proof}. 

Consider that a training dataset $\{\mathbf{g}_i\}_1^{2J}$ that contains $2J$ independent measured images with the first $J$ images satisfying the $H_1$ hypothesis and the rest satisfying the $H_0$ hypothesis is employed to train the multi-task CNN. 
Let $y\in\{0,1\}$ denote the class label, where $y=0$ and $y = 1$ correspond to the hypothesis $H_0$ and $H_1$, respectively. The corresponding class labels and target parameter vectors for the signal present cases are denoted as $\{y_i\}_1^{2J}$ and $\{\bm{\theta}_i\}_1^J$, respectively.
Let the vector $\mathbf{w}=[\mathbf{w_1},\mathbf{w_2}]$ denote the trainable parameters of the multi-task CNN, where $\mathbf{w_1}$ and $\mathbf{w_2}$ denote the weights of the detection and estmation sub-networks, respectively.
In terms of these quantities, the loss functions employed for training the detection and estimation sub-networks can be expressed as:
\begin{subequations}
	\label{eq:loss1}
	\begin{equation}
	\label{eq:loss_detec}
	\mathcal{L}_{\text{Detection}}(\mathbf{w}_1)=-\sum_{j=1}^{2J} p(y_i|\mathbf{g}_i,\mathbf{w}_1)\log p(y_i|\mathbf{g}_i,\mathbf{w}_1)  , 
	\end{equation}
	\begin{equation}
	\label{eq:loss_est}
	\mathcal{L}_{\text{Estimation}}(\mathbf{w}_2)=-\frac{1}{J}\sum_{j=1}^{J} u(\hat{\bm{\theta}}_j(\mathbf{w}_2),\bm{\theta}_j).   
	\end{equation}   
\end{subequations}
The multi-task CNN can be trained by minimizing the loss functions described in Eqn. (\ref{eq:loss1}) in an alternating fashion. At each iteration of the training process, Eqn. (\ref{eq:loss_est}) is minimized first and Eqn. (\ref{eq:loss_detec}) is minimized subsequently. Additional details relevant to a specific implementation of this procedure are described in Sec. \ref{ssec:multi_task_details}.

\subsubsection{Approximating the utility weighted posterior mean using MCMC techniques}
According to Eqn. (\ref{eq:t_decompose}), the utility weighted posterior mean $U(\mathbf{g})$ in Eqn. (\ref{eq:U_g}) is required when computing $T_I(\mathbf{g})$. For a SKS/BKE task, $U(\mathbf{g})$ can be approximated via Monte Carlo integration as:
\begin{equation}
\label{eq:Ug1}
\hat{U}(\mathbf{g})=\frac{1}{J}\sum_{j=1}^{J} u(\hat{\bm{\theta}}_I(\mathbf{g}),\bm{\theta}^{(j)}),
\end{equation}
where $\bm{\theta}^{(j)}$ is sampled from the posterior distribution $p(\bm{\theta}|\mathbf{g},H_1)$. 
A Markov chain with an initial parameter vector $\bm{\theta}^{(0)}$ and a proposal density $q(\bm{\theta}|\bm{\theta}^{(j)})$ can be constructed to generate $\bm{\theta}^{(j)}$, which is described below.

Given $\bm{\theta}^{(i)}$, a candidate parameter vector $\bm{\theta}^\ast$ is drawn from the proposal density and is accepted to the Markov chain with an acceptance probability:
\begin{equation}
\label{eq:accp}
\begin{aligned}
p_a(\bm{\theta}^\ast|&\bm{\theta^{(i)}},\mathbf{g})=\\&\min \left[ 1,\frac{p(\mathbf{g}|\mathbf{s}(\bm{\theta}^\ast),H_1)p(\bm{\theta}^\ast)q(\bm{\theta}^\ast|\bm{\theta}^{(i)})}{p(\mathbf{g}|\mathbf{s}(\bm{\theta}^{(i)}),H_1)p(\bm{\theta}^{(i)})q(\bm{\theta}^{(i)}|\bm{\theta}^\ast)}  \right],
\end{aligned}
\end{equation}
where $p(\bm{\theta})$ is the distribution of the parameter vector. The signal images $\mathbf{s}(\bm{\theta}^{(i)})$ and $\mathbf{s}(\bm{\theta}^\ast)$ are specified by $\bm{\theta}^{(i)}$ and $\mathbf{s}(\bm{\theta}^\ast)$, respectively. The vector $\bm{\theta}^{(i+1)}\equiv\bm{\theta}^\ast$ if the candidate is accepted; otherwise $\bm{\theta}^{(i+1)}\equiv\bm{\theta}^{(i)}$. When a random walk Metropolis-Hastings (RWMH) \cite{pereyra2015survey} is employed, the proposal density $q(\bm{\theta}^\ast|\bm{\theta}^{(i)})$ is a simple Gaussian density: $q(\bm{\theta}^\ast|\bm{\theta}^{(i)})\propto\exp[-\frac{1}{2}(\bm{\theta}^\ast-\bm{\theta}^{(i)})^T\mathbf{K}^{-1}(\bm{\theta}^\ast-\bm{\theta}^{(i)})]$. 

When background variability is considered, $U(\mathbf{g})$ can still be approximated according to Eqn. (\ref{eq:Ug1}) if it is assumed that the background can be described by a stochastic object model (SOM) 
that is parameterized by the parameter vector $\bm{\alpha}$, i.e., $\mathbf{b}\equiv\mathbf{b}(\bm{\alpha})$. The corresponding acceptance rate is:
\begin{equation}
\begin{aligned}
\label{eq:accp_bks}
p_a(\bm{\theta}^\ast,\mathbf{b}(\bm{\alpha}^\ast)|\bm{\theta^{(i)}},\mathbf{b}&(\bm{\alpha}^{(i)}),\mathbf{g})=\\ &\min \left[1, \frac{p_a(\bm{\theta}^\ast,\mathbf{b}(\bm{\alpha}^\ast))}{p_a(\bm{\theta^{(i)}},\mathbf{b}(\bm{\alpha}^{(i)}))} \right],
\end{aligned}
\end{equation}
where
\begin{subequations}
	\begin{equation}
	\begin{aligned}
	p_a(\bm{\theta}^\ast,\mathbf{b}(\bm{\alpha}^\ast)) = &p(\mathbf{g}|\mathbf{s}(\bm{\theta}^\ast),\mathbf{b}(\bm{\alpha}^\ast),H_1)p(\bm{\theta}^\ast)p(\bm{\alpha}^\ast)\\& q_1(\bm{\theta}^\ast|\bm{\theta}^{(i)})q_2(\bm{\alpha}^\ast|\bm{\alpha}^{(i)}), 
	\end{aligned}
	\end{equation}
	\begin{equation}
	\begin{aligned}
    p_a(\bm{\theta^{(i)}},\mathbf{b}(\bm{\alpha}^{(i)})) = &p(\mathbf{g}|\mathbf{s}(\bm{\theta}^{(i)}),\mathbf{b}(\bm{\alpha}^{(i)}),H_1)p(\bm{\theta}^{(i)})\\&p(\bm{\alpha}^{(i)}) q_1(\bm{\theta}^{(i)}|\bm{\theta}^\ast)q_2(\bm{\alpha}^{(i)}|\bm{\alpha}^\ast).
    \end{aligned}
	\end{equation}   
\end{subequations}
Here, $\mathbf{b}(\bm{\alpha}^{(i)})$ and $\mathbf{b}(\bm{\alpha}^\ast)$ are the background images determined by the parameter vectors $\bm{\alpha}^{(i)}$ and $\bm{\alpha}^\ast$, respectively.
The quantity $p(\hat{\bm{\alpha}})$ is the corresponding probability density, and $q_1(\cdot)$ and $q_2(\cdot)$ are the proposal densities for the signal and background parameters, respectively. The pair $(\bm{\theta}^{(i)},\bm{\alpha}^{(i)})$ are sampled from the distribution $p(\bm{\theta},\mathbf{b}(\bm{\alpha})|\mathbf{g},H_1)$, and $(\bm{\theta}^\ast,\bm{\alpha}^\ast)$ are candidate parameter vectors that are drawn from the proposal densities. A Markov chain for generating $(\bm{\theta}^{(i)},\bm{\alpha}^{(i)})$ can then be established in a similar way as described above.

\subsection{Supervised learning-based sub-ideal NO}
As described above, the proposed hybrid method can approximate the IO for detection-estimation tasks. However, the MCMC technique is a component of the hybrid method, which requires knowledge of background density $p(\bm{\alpha})$ if background variability is considered. 
This currently limits the application of the hybrid method to certain object models \cite{zhou2019approximating}. To circumvent this, a purely supervised learning-based sub-optimal NO can be employed.

\begin{figure}[ht]
	\centering
		\includegraphics[width=0.5\textwidth]{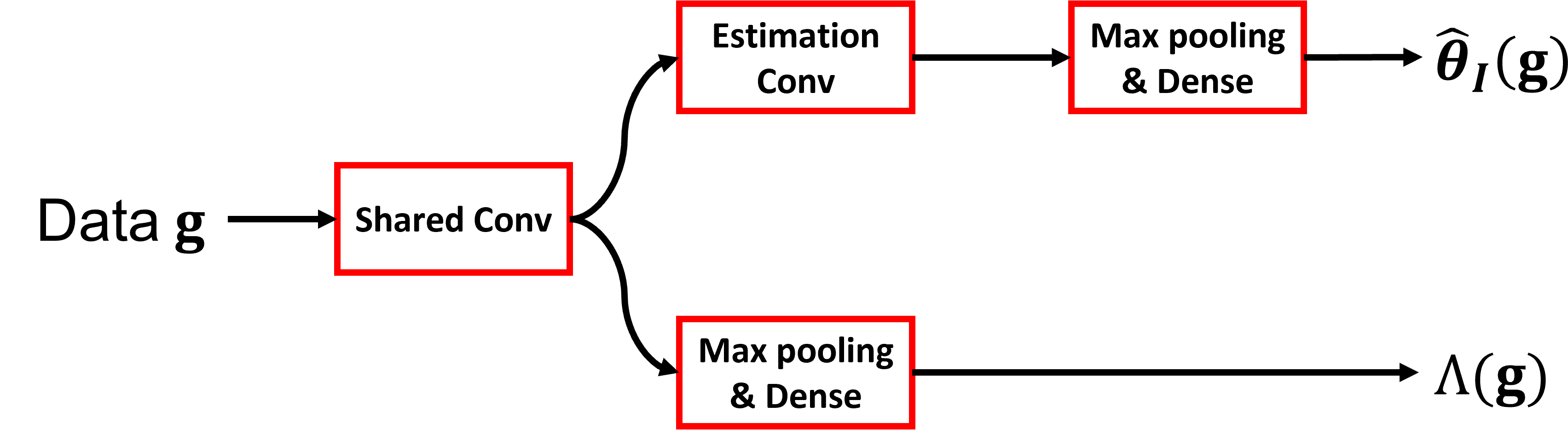}
	\caption{A schematic of the multi-task CNN employed to form the sub-ideal NO. The Shared Conv and Estimation Conv are two blocks that comprise several convolutional layers. The convolutional layers in the Shared Conv are shared by the detection and estimation sub-networks  (i.e., the bottom and top network branches). The layers in the Estimation Conv are employed by the estimation sub-network only.}
	\vspace{-0.1in}
    \label{fig:sub_architecture}
\end{figure}

A sub-optimal NO that is established by use of only supervised-learning
can be readily obtained by eliminating 
 the MCMC method from the hybrid method as depicted in Fig. \ref{fig:sub_architecture}. This modification removes the influence of the estimation result on the detection performance. The corresponding test statistic and estimator are given by:
\begin{subequations}
\label{eq:Sub_test_estimate}
\begin{equation}
T_{Sub}(\mathbf{g})=\frac{p(\mathbf{g}|H_1)}{p(\mathbf{g}|H_0)}=\Lambda(\mathbf{g})
\end{equation}

\begin{equation}
\hat{\bm{\theta}}_{Sub}(\mathbf{g})=\argmax_{\hat{\bm{\theta}}} \left\{\int
p(\bm{\theta})\Lambda(\mathbf{g}|\bm{\theta})u(\bm{\theta},\hat{\bm{\theta}})d\bm{\theta}\right\},
\end{equation}
 \end{subequations}
where $T_{Sub}(\mathbf{g})$ is IO test statistic for binary signal detection tasks and $\hat{\bm{\theta}}_{Sub}(\mathbf{g})$ IO estimator for detection-estimation tasks.

\section{Numerical studies}
\label{sec:numerical_studies}
Computer-simulation studies were performed to investigate the proposed NOs for detection-estimation tasks.
The considered signal detection-estimation tasks included both BKE and BKS tasks. A lumpy background (LB) model \cite{kupinski2003ideal} and a clustered lumpy background (CLB) model \cite{bochud1999statistical} were employed in the BKS tasks.

The imaging system considered was an idealized parallel-hole
collimator system that was specified by a linear C-D mapping with Gaussian point response functions (PRFs)~\cite{kupinski2003ideal}: 
\begin{equation}
h_m(\mathbf{r})=\frac{h}{2\pi w_{m}^2}\exp\left[ -\frac{(\mathbf{r-r}_m)^T(\mathbf{r-r}_m)}{2w_{m}^2} \right], 
\label{eq:parallel-hole}
\end{equation}
where $h$ and $w_m$ are the height and width of the PRFs, respectively. 

The signal to be detected and estimated was modeled by a 2D Gaussian function:
\begin{equation}
f_s(\mathbf{r})={A_s}\exp\left[ -\frac{(\mathbf{r-r}_s)^T(\mathbf{r-r}_s)}{2w_{s}^2} \right], 
\label{eq:sig_fun}
\end{equation}
where $A_{s}$ is the signal amplitude, $w_s$ is the signal width and $\mathbf{r_s}$ is the center of signal. These signal parameters can be random when a detection-estimation task is specified. Considering the specified imaging system in Eqn. (\ref{eq:parallel-hole}), the $m^{th}$ element $[\mathbf{s}]_m$ of the signal image $\mathbf{s}$ is given by:
\begin{equation}
[\mathbf{s}]_m= \frac{{A_s}{h}{w_s^2}}{w_m^2+w_s^2}\exp\left[ -\frac{(\mathbf{r}_m-\mathbf{r}_s)^T(\mathbf{r}_m-\mathbf{r}_s)}{2(w_{m}^2+w_{s}^2)}\right].
\label{eq:sig_bks}
\end{equation}

In the studies described below, the AEROC was employed to quantify observer performance. The AEROC was
estimated by use of a nonparametric estimator \cite{wunderlich2014nonparametric} and the uncertainty
in the estimates was conveyed by use of a 90\% confidence interval.

\subsection{SKS/BKE signal detection-estimation task}
\label{ssec:numerical_bke}
In the SKS/BKE case considered, the task was to detect a random signal and estimate its amplitude. This task can be viewed as a surrogate for tumor detection in positron emission tomography (PET) \cite{rezai2013radiologist}. 
In this task, the dimensions of $\mathbf{g}$, $\mathbf{b}$, and $\mathbf{n}$ were $64\times64$ pixels. Without loss of generality, $\mathbf{b=0}$. The signal defined in Eqn.~(\ref{eq:sig_fun}) was employed with $w_s=1$ and $\mathbf{r}_s=[32;32]^T$. The to-be-estimated signal amplitude was sampled from a Gaussian distribution with mean $\mu_A=9$ and standard deviation $\sigma_A=4$, i.e., $A_s\sim\mathcal{N}(9,4^2)$. The assumed parameters of the imaging system defined in Eqn. (\ref{eq:parallel-hole}) were $h=16$ and $w_m=3.87$. The standard deviation $\sigma_n$ of Gaussian noise was set to 40. To define an EROC curve, a Gaussian utility function, $u(\hat{A_s},A_s)=\exp[-(\hat{A_s}-A_s)^2/(2\sigma_u^2)]$, was employed where $\sigma_u=3$. 

Wunderlich \etal \cite{wunderlich2016optimal} provided the optimal decision rule for this case that is expressed as:
\begin{equation}
\label{eq:analy}
\begin{aligned}
&\hat{A}_I(\mathbf{g})=\frac{\sigma_A^2\mathbf{s}_{ref}^T\mathbf{g}+\sigma_n^2\mu_A}{\sigma_n^2+\sigma_A^2\mathbf{s}_{ref}^T\mathbf{s}_{ref}}\\
&T_I(\mathbf{g})=\mu_A(\mathbf{s}_{ref}^T\mathbf{g})+\frac{\sigma_A^2}{2\sigma_n^2}(\mathbf{s}_{ref}^T\mathbf{g})^2.
\end{aligned}
\end{equation}
Here, $\mathbf{s}_{ref}$ is a reference signal whose components are defined as
\begin{equation}
[\mathbf{s}_{ref}]_m= \frac{{h}{w_s^2}}{w_m^2+w_s^2}\exp\left[ -\frac{(\mathbf{r}_m-\mathbf{r}_s)^T(\mathbf{r}_m-\mathbf{r}_s)}{2(w_{m}^2+w_{s}^2)}\right].
\label{eq:sig_ref}
\end{equation}
This analytic IO decision strategy was implemented as a reference method that
the proposed hybrid method and sub-ideal NO were compared against. The Gaussian proposal density in the MCMC method that was employed to estimate $U(\mathbf{g})$ was chosen to have a standard deviation of 3.

\subsection{SKS/BKS signal detection-estimation tasks with a lumpy background model}
\label{ssec:numerical_lb}
In the first BKS task considered, the Gaussian signal defined in Eqn.~(\ref{eq:sig_fun}) was employed with $A_s=6$ and $w_s=3$. The signal location $\mathbf{r_s}$ was a two-dimensional random vector whose components were independently sampled from a uniform distribution on the interval $(16,48)$.  A quadratic utility function $u_1(\mathbf{\hat{r}_s},\mathbf{r_s})=1-\frac{1}{\epsilon_1}\Vert\mathbf{\hat{r}_s}-\mathbf{r_s}\Vert_2^2$ and an $l^1$-norm based utility function  $u_2(\mathbf{\hat{r}_s},\mathbf{r_s})=1-\frac{1}{\epsilon_2}\Vert\mathbf{\hat{r}_s}-\mathbf{r_s}\Vert_1$ were employed, where
$\epsilon_1=100$ and 200, and $\epsilon_2=20$, respectively. A quadratic utility function focuses on large errors while an $l^1$-norm based one emphasizes more on small errors. The different values $\epsilon_1$ were employed to investigate the influences of estimation tasks on the test statistics according to Eqn. (\ref{eq:t_decompose}).
Compared with previously investigated detection-localization tasks \cite{khurd2005decision,zhou2020approximating}, the task in this study can be considered as a straightforward generalization.
 
\begin{figure}[ht]
	\centering
	\subfigure[]
		{\includegraphics[width=0.1571\textwidth]{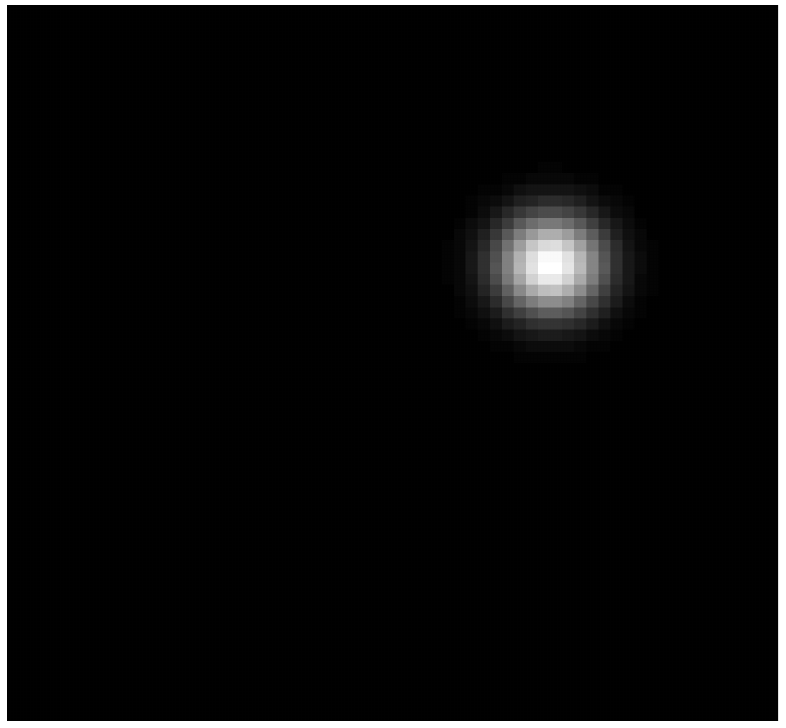}}
	\subfigure[]
		{\includegraphics[width=0.157\textwidth]{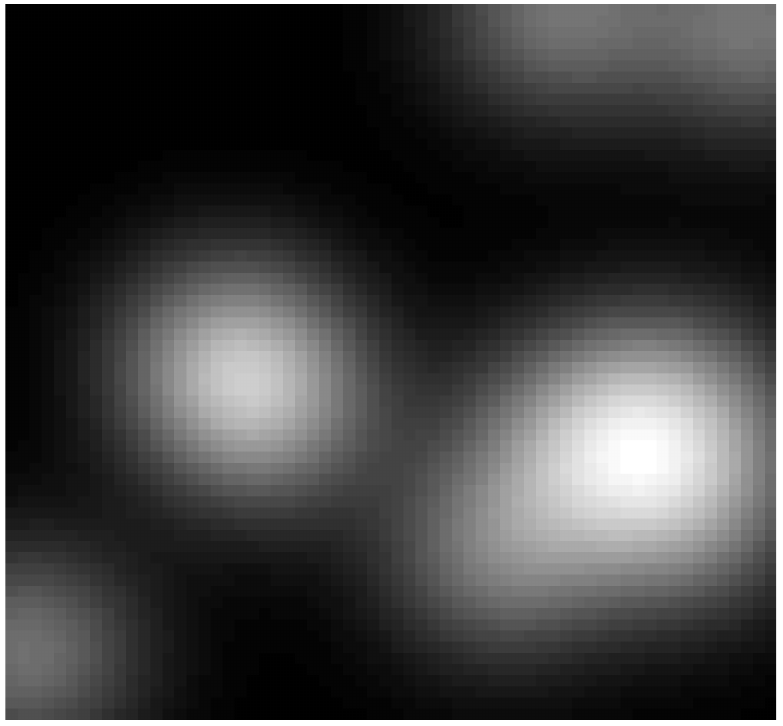}}
	\subfigure[]
	{\includegraphics[width=0.1571\textwidth]{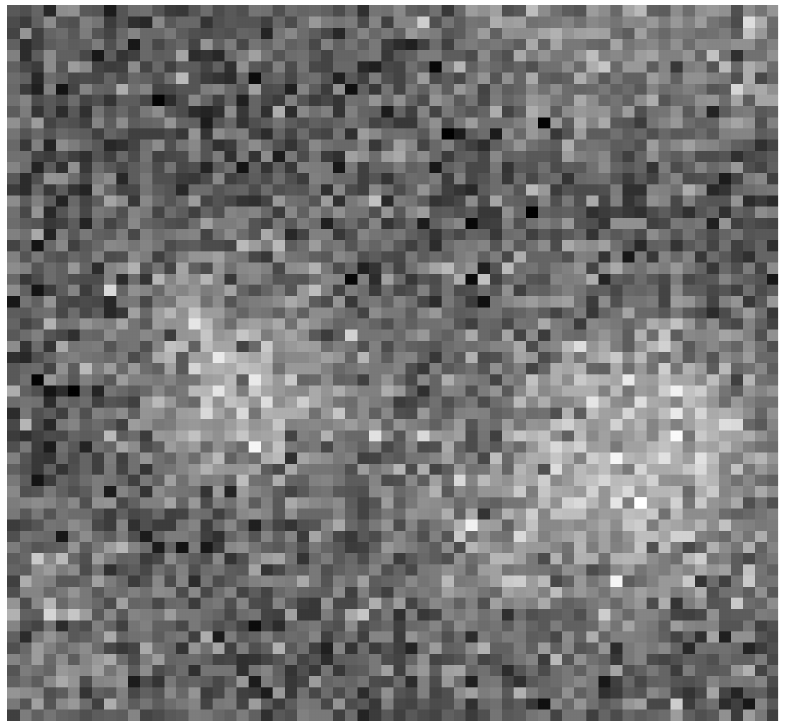}}
\vspace{-0.1in}
	\caption{Realizations that depict (a) a possible signal $\mathbf{s}$, (b) a background image $\mathbf{b}$, and (c) the corresponding noisy measurement $\mathbf{g}$ for the SKS/BKS signal detection-estimation tasks with a LB model. The dimensions of the images were $64\times64$.}
	\label{fig:sample_lumpy}
\end{figure}

To emulate background variability, a lumpy object model \cite{kupinski2005small} was utilized:
\begin{equation}
f_b(\mathbf{r})=\sum^{N_b}_{n=1}l(\mathbf{r-r}_n|a,w_b),
\label{eq:lumpy}
\end{equation}
where $N_b\sim P(\bar{N})$ denotes the number of the lumps with 
$P(\bar{N})$ denoting a Poisson distribution with the mean $\bar{N}=5$.
The lump function $l(\mathbf{r-r}_n|a,w_b)$ was modeled by a 2D Gaussian function with lump amplitude $a=10$ and lump width $w_b=7$:
\begin{equation}
l(\mathbf{r-r}_n|a,w_b)=a\exp(-\frac{(\mathbf{r-r}_n)^T(\mathbf{r-r}_n)}{2w^2_b}). 
\end{equation}
Here, $\mathbf{r}_n$ denotes the center location of the $n^{th}$ lump 
that was sampled from a uniform distribution over the spatial support of the image. The dimensions of $\mathbf{s}$, $\mathbf{b}$, $\mathbf{n}$ and $\mathbf{g}$ in Eqn.~(\ref{eq:hypo}) were $64\times64$. The imaging system was specified by $h=40$ and $w_m=0.5$. 
Given the assumed imaging system, the $m^{th}$ element $[\mathbf{b}]_m$ of the background image $\mathbf{b}$ is given by:
\begin{equation}
[\mathbf{b}]_m= \frac{{a}{h}{w_b^2}}{w_m^2+w_b^2}\sum_{n=1}^{N_b}\exp\left[ -\frac{(\mathbf{r}_n-\mathbf{r}_m)^T(\mathbf{r}_n-\mathbf{r}_m)}{2(w_{m}^2+w_{b}^2)}\right].
\label{eq:lumpy_bks}
\end{equation}
The measurement noise  was described by i.i.d. Gaussian random variables with a mean of 0 and a standard deviation of 320. One realization of the signal image $\mathbf{s}$, the background image $\mathbf{b}$, and the corresponding signal-present noisy measurement $\mathbf{g}$ are shown in Figure \ref{fig:sample_lumpy}.

Because the IO decision rule in this case cannot be analytically computed, an MCMC-based IO approximation method was established as a reference method. Unlike the proposed hybrid method, this MCMC-based method requires a strong constraint on the utility function. The details of the MCMC approximated IO (MCMC-IO) are provided in Appendix \ref{ap:MCMC-IO}. The performances of the proposed hybrid method and sub-ideal NO were compared to that of MCMC-IO when the quadratic utility function was employed. When the $l^1$-norm based utility function was considered, the MCMC-IO cannot be employed and the SLO described in Sec. \ref{ssec:slo} was utilized as the reference observer. {To implement the SLO, the covariance matrix was estimated by use of the covariance matrix decomposition method~\cite{barrett2013foundations,lee2013assessment} and 4,000 signal-present and 4,000 signal-absent noiseless images were utilized.} The Gaussian proposal density in the MCMC method that was employed to estimate $U(\mathbf{g})$ was chosen to have a standard deviation of 4 for each location coordinate.
\vspace{-0.1in}

\subsection{SKS/BKS signal detection-estimation tasks with a clustered lumpy background model}
In the second BKS task considered, the signal image $\mathbf{s}$ had an amplitude of 0.05 and signal location $\mathbf{r}_s=[32;32]^T$. The width of $\mathbf{s}$ was a random vector and was sampled from a uniform distribution on the interval $(1,6)$.
A Gaussian utility function, $u(\hat{w},w)=\exp[-(\hat{w}-w)^2/(2\sigma_u^2)]$, was employed with $\sigma_u=3$.

A clustered lumpy background (CLB) model \cite{bochud1999statistical} was employed to emulate background variability. The $m^{th}$ element $[\mathbf{b}]_m$ of the background image $\mathbf{b}$ was computed as \cite{bochud1999statistical}:
\begin{equation}
[\mathbf{b}]_m=\sum_{k=1}^K\sum_{n=1}^{N_k}l(\mathbf{r}_m-\mathbf{r}_k-\mathbf{r}_{kn}|\mathbf{R}_{\theta_{kn}}).
\label{eq:clb_bks}
\end{equation}
Here, $K\sim P(\bar{K})$ denotes the number of the clusters with 
$P(\bar{K})$ denoting a Poisson distribution with the mean $\bar{K}$, $N_k$ denotes the number of blobs in the $k^{th}$ cluster that was sampled from a Poisson distribution with the mean of $\bar{N}$: $N_k\sim P(\bar{N})$, $\mathbf{r}_k$ denotes the center location of the $k^{th}$ cluster that was sampled from a uniform distribution over the spatial support of the image, and $\mathbf{r}_{kn}$ denotes the center location of the $n^{th}$ blob in the $k^{th}$ cluster that was sampled from a Gaussian distribution with the center of $\mathbf{r}_k$ and standard deviation of $\sigma$. The blob function $l(\mathbf{r}|\mathbf{R}_{\theta_{kn}})$ was specified as:
\begin{equation}
l(\mathbf{r}|\mathbf{R}_{\theta_{kn}})=\exp\left(-\alpha\frac{\Vert\mathbf{R}_{\theta_{kn}}\mathbf{r}\Vert^\beta}{L(\mathbf{R}_{\theta_{kn}}\mathbf{r})}\right).
\label{eq:clb_blob}
\end{equation}
Here, $L(\mathbf{r})$ is computed as the “radius” of the ellipse with half-axes $L_x$ and $L_y$, and $\mathbf{R}_{\theta_{kn}}$ is the rotation matrix corresponding to the angle $\theta_{kn}$ that was sampled uniformly between 0 and $2\pi$. The generated background images were normalized to the range between 0 and 1. The parameters that specify the CLB model employed in this study are summarized in Table. \ref{tab:clb}.
\begin{table}[ht]
	\centering
	\caption{Parameters for generating CLB images}
	\begin{tabular}{cc c c c c c }
    \hline\hline
    $\bar{K}$ &$\bar{N}$ &$L_x$ &$L_y$ &$\alpha$ &$\beta$ &$\sigma$  \\
    
    70        &20        &5     &2     &2.1      &0.5     &12       \\
	\hline\hline
	\end{tabular}
	\label{tab:clb}
\end{table}

\begin{figure}[ht]
	\centering
	\subfigure[]
		{\includegraphics[width=0.1555\textwidth]{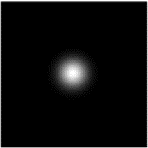}}
	\subfigure[]
		{\includegraphics[width=0.1555\textwidth]{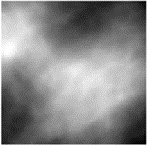}}
	\subfigure[]
	{\includegraphics[width=0.1635\textwidth]{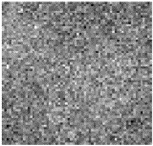}}
	\vspace{-0.1in}
	\caption{Images that depict (a) a possible signal $\mathbf{s}$, (b) a background image $\mathbf{b}$, and (c) the corresponding noisy measurement $\mathbf{g}$ for the SKS/BKS signal detection-estimation tasks with a CLB model. The dimensions of the images were $64\times64$.}
	\label{fig:sample_clb}
\end{figure}

The measurement noise was described by i.i.d. Gaussian random variables with a mean of 0 and a standard deviation of 0.33. One realization of the signal image $\mathbf{s}$, the background image $\mathbf{b}$, and the corresponding signal-present noisy measurement $\mathbf{g}$ are shown in Figure \ref{fig:sample_clb}.

\vspace{-0.1in}
\vspace{0.1in}
The SLO described in Sec. \ref{ssec:slo} was utilized as the reference observer. The details regarding the implementation are described in Sec. \ref{ssec:numerical_bke}.

\subsection{Multi-task CNN training details}
\label{ssec:multi_task_details}
Details regarding the implementation of multi-task CNN described in Sec. \ref{sssec:multi-task} are described here. Each convolutional layer in the Shared Conv block comprised 64 filters with $5\times 5$ spatial support followed by a Leaky ReLU activation function \cite{springenberg2014striving}. 
In the detection sub-network, a max-pooling layer \cite{scherer2010evaluation} was employed to subsample the feature maps and the last layer was fully connected (FC) and employed a sigmoid activation function for estimation of the posterior probability.  
In the estimation sub-network, the Estimation Conv block included additional convolutional layers to account for the fact that estimation can be a more complicated task. The architecture for both blocks were identical. A max-pooling layer and a FC layer were added to compute the estimate.

The train-validation-test scheme \cite{goodfellow2016deep} was employed to train and evaluate multi-task CNNs. 
The initial training dataset included 150,000 signal-present images and 150,000 signal-absent images for the BKS detection-estimation task with the lumpy background. For tasks that involved the clustered lumpy background, 200,000 signal-present images and 200,000 signal-absent images were employed.
To mitigate overfitting, a “semi-online learning” method \cite{zhou2019approximating, zhou2020approximating} was employed 
in which measurement noise was generated on-the-fly and added to noiseless images drawn from the finite initial training dataset.
The validation datatset included 1000 signal-present images and 1000 signal absent images.
Finally, the test dataset comprised 1000 signal-present images and 1000 signal-absent images.

The multi-task CNN was trained by minimizing the loss functions described in Eqn. (\ref{eq:loss1}) on mini-batches in an alternating fashion. At each iteration of the training process, Eqn. (\ref{eq:loss_est}) was minimized first and Eqn. (\ref{eq:loss_detec}) was minimized subsequently. Multi-task CNNs comprising different number of layers were trained for 200,000 mini-batches.
Each mini-batch contained 200 signal-absent images and 200 signal-present images randomly selected from the training data set.
The Adam optimizer~\cite{kingma2014adam} with a learning rate of 0.00001 was employed for model training. 

In order to accurately approximate the IO, the CNN architecture needs to possess sufficient capacity \cite{zhou2019approximating}.
For determining the optimal architecture of the multi-task CNN, 
the training process started from an architecture with one convolutional layer in both the Shared Conv and Estimation Conv blocks to formulate baseline detection and estimation sub-networks. More layers were added gradually as described below. 
The optimal number of convolutional layers for the detection sub-network was determined by adding more  layers in the Shared Conv block until the cross-entropy on the validation dataset did not significantly decrease. The loss functions were considered as significantly decreased if its decrements are at least $1.0\%$ of that produced by multi-task CNN with one less convolutional layer. After the number of layers for the detection sub-network was determined, the optimal number of layers for estimation sub-network was specified by increasing the number of convolutional layers in the Estimation Conv block. The training and implementation of the multi-task CNN were performed using Tensorflow~\cite{tensorflow2015-whitepaper}.

\section{Results}
\label{sec:results}
\subsection{SKS/BKE signal detection-estimation tasks}

\begin{figure}[ht]
	\centering
	\includegraphics[width=0.47\textwidth]{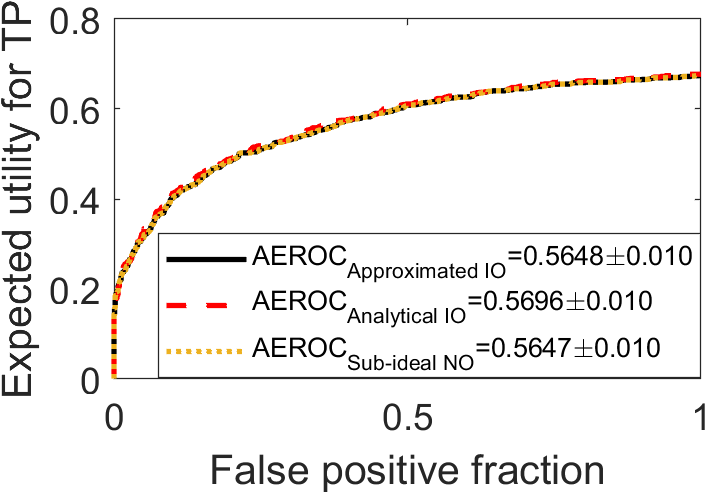}
	\caption{The EROC curves produced by the approximated IO (black), the analytical computation (red), and sub-ideal NO (yellow) for the BKE task were statistically equivalent.
	}
	\label{fig:result_bke}
\end{figure}

The optimal network architecture was determined to contain three convolutional layers in the Shared Conv block and two layers in the Estimation Conv block.
In Figure \ref{fig:result_bke}, the EROC curves produced by the approximated IO (black solid curve) and the sub-ideal NO (yellow dashed curve) are compared with that produced by the analytical computation (red dashed curve). The AEROC values were $0.565\pm0.010$, $0.565\pm0.010$, and $0.570\pm0.010$ corresponding to the approximated IO, the sub-ideal NO, and the analytical computation, respectively. The corresponding EROC curves were statistically equivalent in this SKE/BKE case.
\vspace{-0.2in}

\subsection{SKS/BKS signal detection-estimation tasks with a lumpy background model}
\begin{figure}[ht]
	\centering
	\includegraphics[width=0.47\textwidth]{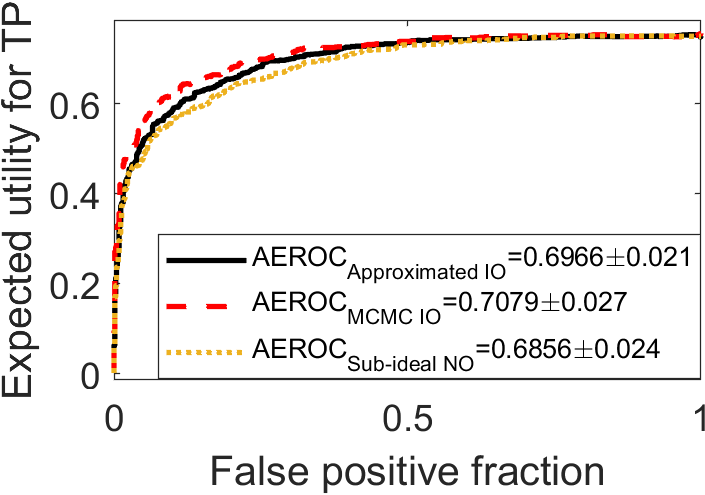}
	\caption{The EROC curves produced by the approximated IO (black), the MCMC-IO (red), and sub-ideal NO (yellow) for the BKS task with the lumpy background model are compared. The corresponding AEROC values were statistically equivalent in this case. The quadratic utility function described in Sec. \ref{ssec:numerical_lb} was employed with $\epsilon_1=200$.
	}
	\vspace{-0.1in}
	\label{fig:result_bks_lb}
\end{figure}

The optimal network architecture was determined to contain seven convolutional layers in the Shared Conv block and three layers in the Estimation Conv block.
To compare the proposed methods with the IO approximated by the MCMC method, a quadratic utility function described in Sec. \ref{ssec:numerical_lb} with $\epsilon_1=200$ was employed. For this case, the EROC curves generated by use of the approximated IO (black solid curve) and the sub-ideal NO (yellow dashed curve) are compared with that produced by the MCMC method (red dashed curve) in Figure \ref{fig:result_bks_lb}. The AEROC values were $0.697\pm0.021$, $0.686\pm0.024$, and $0.708\pm0.027$ corresponding to the approximated IO, the sub-ideal NO, and the MCMC-IO, respectively. The corresponding EROC curves were statistically equivalent in this BKS signal detection-estimation task.

\begin{figure}[ht]
	\centering
	\includegraphics[width=0.47\textwidth]{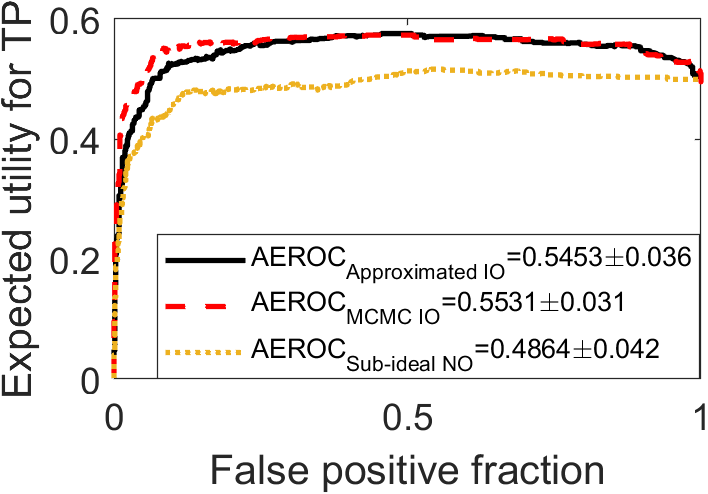}
	\caption{The EROC curves produced by the approximated IO (black), the MCMC-IO (red), and the sub-ideal NO (yellow) for the BKS task with the lumpy background model are compared. The quadratic utility function described in Sec. \ref{ssec:numerical_lb} was employed with $\epsilon_1=100$. 
	}
	\vspace{-0.1in}
	\label{fig:result_bks_lb_l2_100}
\end{figure}

A quadratic utility function described in Sec. \ref{ssec:numerical_lb} with $\epsilon_1=100$ was employed to investigate a situation where the performance of sub-ideal NO is worse than that of the approximated IO. Compared with the case addressed in Figure \ref{fig:result_bks_lb}, the variance of $U(\mathbf{g})$ increased from 0.1708 to 0.7295, indicating that the estimation performance has larger influence on the  test statistic according to Eqn. (\ref{eq:t_decompose}). Because of this increase in variance,  the performance of the hybrid approximated IO became better than that of the sub-ideal NO.
In Figure \ref{fig:result_bks_lb_l2_100}, the EROC curves generated by use of the approximated IO (black solid curve) and the sub-ideal NO (yellow dashed curve) are compared with that produced by the MCMC method (red dashed curve). The AEROC values were $0.545\pm0.036$, $0.486\pm0.042$, and $0.553\pm0.031$ corresponding to the approximated IO, the sub-ideal NO, and the MCMC-IO, respectively. The EROC curves corresponding to the approximated IO and the MCMC-IO are in close agreement in this SKS/BKS signal detection-estimation task, while the difference between the approximated IO and the sub-ideal NO is statistically significant. 

Intuition for this result can be gained by noting that, when the variance of $U(\mathbf{g})$ is small, Eqn. (\ref{eq:t_decompose}) can be approximated as $T_I(\mathbf{g})\approx \Lambda(\mathbf{g}) \langle {U}(\mathbf{g}) \rangle $, where $\langle {U}(\mathbf{g}) \rangle$ is the mean
of $U(\mathbf{g})$. In this case, because their test statistics differ by (approximately) a constant, the hybrid approximated IO and sub-ideal NO will perform similarly. However, this is not to be expected when the variance of $U(\mathbf{g})$ is large because the test statistics $T_I(\mathbf{g})$ and $\Lambda(\mathbf{g})$ will not simply differ by  a constant and they will not generally be related by a monotonic transformation.

\begin{figure}[ht]
	\centering
	\includegraphics[width=0.47\textwidth]{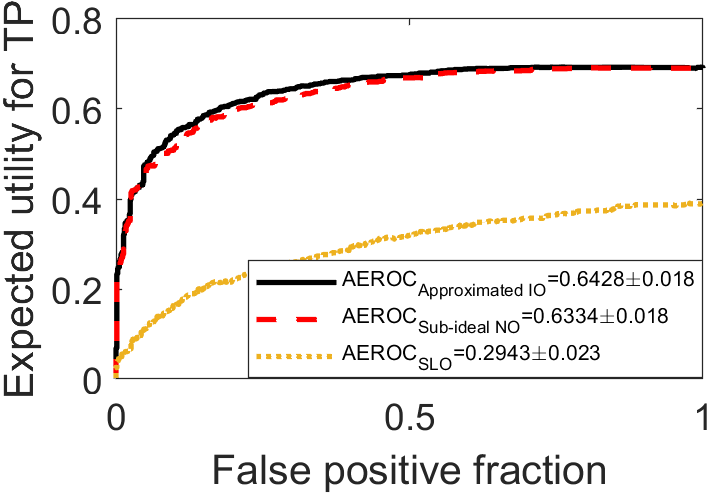}
	\caption{The EROC curves produced by the approximated IO (black), the sub-ideal NO (red), and the SLO (yellow) for the BKS task with the lumpy background model are compared. The $l^1$-norm based utility function described in Sec. \ref{ssec:numerical_lb} was employed with $\epsilon_2=20$. 
	}
	\vspace{-0.1in}
	\label{fig:result_bks_lb_l1}
\end{figure}

To demonstrate a case where the MCMC-IO cannot be applied but the hybrid method can, the $l^1$-norm based utility function described in Sec. \ref{ssec:numerical_lb} was employed. In this case, the SLO was employed as the reference observer. Compared with a quadratic utility function, small errors weight more in an $l^1$-norm based one. As shown in Figure \ref{fig:result_bks_lb_l1}, the EROC curves generated by use of the approximated IO (black solid curve) and the sub-ideal NO (red dashed curve) are compared with that produced by the SLO (yellow dashed curve). The AEROC values
corresponding to the approximated IO, the sub-ideal NO, and the SLO 
were $0.643\pm0.018$, $0.633\pm0.018$, and $0.294\pm0.023$, respectively. 
The EROC curves corresponding to the approximated IO and the sub-ideal NO were in close agreement in this task. As expected, the AEROC value produced by the SLO was significantly smaller. The results demonstrated that the performance of sub-ideal NO was significantly better than that of the SLO.
\vspace{-0.1in}

\subsection{BKS signal detection-estimation tasks with a CLB model}
\begin{figure}[ht]
	\centering
	\includegraphics[width=0.47\textwidth]{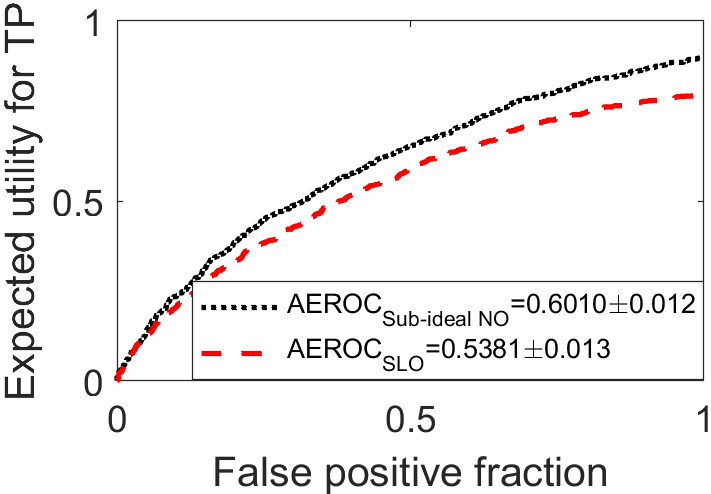}
	\caption{The EROC curves corresponding to the sub-ideal NO (black) are compared with that generated by SLO (red) for the BKS task with the clustered lumpy background model. The AEROC value produced by the sub-ideal NO was larger than that produced by the SLO.
	}
	\vspace{-0.1in}
	\label{fig:result_bks_Clb}
\end{figure}

To investigate a case where only the sub-ideal purely supervised NO could be applied, the task involving the CLB background model was employed. The optimal network architecture was determined to contain eight convolutional layers in the Shared Conv block and three layers in the Estimation Conv block.
Considering that current MCMC methods have not been applied to the CLB object model, the hybrid method was not employed and the SLO was employed as the reference method.
The EROC curves corresponding to the sub-ideal NO (black dashed curve) are compared with that generated by SLO (red dashed curve) in Figure \ref{fig:result_bks_Clb}. The AEROC value corresponding to the sub-ideal purely supervised NO was $0.601\pm0.012$, which was larger than the $0.538\pm0.013$ produced by the SLO.
\vspace{-0.1in}

\section{Summary}
\label{sec:discussion}
General signal detection-estimation tasks are frequently considered in medical imaging.
For detection-estimation tasks, the EROC  curve has been
proposed for evaluating the performance of observers. However, in practice, it 
is difficult to accurately approximate the IO that maximizes the
AEROC for a general detection-estimation task.
In this work, a hybrid approach was developed that combines a multi-task CNN and a MCMC method in order to approximate the IO for detection-estimation tasks. Compared with the MCMC-IO, the hybrid method is not limited to use of specific utility functions. Additionally, a supervised learning-based sub-ideal NO was designed for signal detection-estimation tasks. 
Both SKS/BKE and SKS/BKS tasks were considered and computer-simulation studies were conducted to validate the proposed methods. The proposed hybrid method provides a new approach for approximating the IO and may advance the  application  of  EROC  analysis  for  optimizing  imaging systems.

The proposed methods possess certain limitations. The MCMC methods are employed in the hybrid framework for IO approximation, which makes the limitations of MCMC methods also exist in the hybrid methods. Numerous practical issues such as the design of proposal densities from which the Markov chain can be efficiently generated need to be addressed. Because of this, the hybrid method has been limited to relatively simple object models. An advanced method called MCMC-GAN \cite{zhou2020markov} can potentially solve this problem. By replacing the classic MCMC methods in the hybrid method with MCMC-GAN, the IO approximation method can be extended to more complicated background models.
In addition, the supervised learning-based method may require a large amount of training data to accurately train multi-task CNNs. To address this limitation, one may establish a stochastic object model (SOM) from experimental data by training an AmbientGAN \cite{bora2018ambientgan,zhou2020progressively}. Given a well-established SOM, one can produce large amount of training samples to train the  multi-task CNNs. 

There remains several other topics for future investigation. 
The proposed methods should be further validated by use of a variety of
image data and detection-estimation tasks that  address  real-world problems.
Additionally, it will be important to quantify the effect of insufficient training data on the proposed methods.
Finally, it will be interesting to employ the proposed methods to perform task-based performance evaluation of deep learning-based image restoration  \cite{kelkar2021task,li2021task,li2021assessing,you2018structurally} and image reconstruction \cite{kelkar2021compressible,kelly2017deep,wang2020deep} techniques.

\begin{appendices}

\section{Approximating the ideal estimate using multi-task CNNs}
\label{ap:est_proof}
According to Eqn. (\ref{eq:IO_test_estimate_est}), the ideal estimate can be defined as:
\begin{equation}
\label{eq:appendix_theta}
\hat{\bm{\theta}}_I(\mathbf{g})=\Lambda(\mathbf{g})\left\{\argmax_{\hat{\bm{\theta}}} \int pr(\bm{\theta}|\mathbf{g},H_1)u(\hat{\bm{\theta}},\bm{\theta})d\bm{\theta}\right\},
\end{equation}
where $\Lambda(\mathbf{g})$ is the likelihood ratio. Considering that $\Lambda(\mathbf{g})$ is independent to $\hat{\bm{\theta}}$, Eqn. (\ref{eq:appendix_theta}) can be written as:
\begin{equation}
\label{eq:appendix_theta2}
\hat{\bm{\theta}}_I(\mathbf{g})=\argmax_{\hat{\bm{\theta}}} \int pr(\bm{\theta}|\mathbf{g},H_1)u(\hat{\bm{\theta}},\bm{\theta})d\bm{\theta}=\argmax_{\hat{\bm{\theta}}} U(\mathbf{g}).
\end{equation}
A supervised learning-based method can be employed to approximate $\hat{\bm{\theta}}_I(\mathbf{g})$ that maximizes the expectation $E[U(\mathbf{g})]$ over an ensemble of training data, where the expectation is taken over $\mathbf{g}$. Assuming $\mathbf{g}$ is signal-present $p(\mathbf{g})=p(\mathbf{g}|H_1)$, $E[U(\mathbf{g})]$ can be depicted as:
\begin{equation}
\label{eq:appendix_theta3}
\begin{aligned}
E[U(\mathbf{g})]&=\iint pr(\bm{\theta}|\mathbf{g},H_1)u(\hat{\bm{\theta}},\bm{\theta})p(\mathbf{g}|H_1)d\bm{\theta}d\mathbf{g}\\
&=\iint pr(\bm{\theta},\mathbf{g}|H_1)u(\hat{\bm{\theta}},\bm{\theta})d\bm{\theta}d\mathbf{g}=E[u(\hat{\bm{\theta}},\bm{\theta})],
\end{aligned}
\end{equation}
where the expectation is taken over the random variables $\mathbf{g}$, $\mathbf{\hat{\bm{\theta}}}$, and $\bm{\theta}$. Considering a training dataset $\{\mathbf{g}_j,\bm{\theta}_j\}_1^{J}$ that contains $J$ independent signal-present images with corresponding parameter vectors for joint tasks. Similar to Wunderlich \etal \cite{wunderlich2014nonparametric}, $E[u(\hat{\bm{\theta}},\bm{\theta})]$ can be estimated with the unbiased estimators:
\begin{equation}
\label{eq:appendix_theta2}
\hat{E}[u(\hat{\bm{\theta}},\bm{\theta})]=\frac{1}{J}\sum_{j=1}^J u(\bm{\theta}_j,\hat{\bm{\theta}}_j),
\end{equation}
where $\hat{\bm{\theta}}_j$ is the observer's estimate of $\bm{\theta}_j$. Let $\mathbf{w}_2$ denotes the weight vector that parameterizes an estimation sub-network and $\hat{\bm{\theta}}(\mathbf{w}_2)$ is the signal parameter vector estimated by the CNN. The desired weight vector $\mathbf{w}_{I}$ for the estimation sub-network can be determined by optimizing the following loss function:
\begin{equation}
	\label{eq:appendix_loss_est}
	\mathbf{w}_{I}=\argmin_{\mathbf{w}_{2}}\frac{1}{J}\sum_{j=1}^{J} -u(\hat{\bm{\theta}}(\mathbf{w}_2),\bm{\theta}),    
	\end{equation} 
\end{appendices}
which leads to the loss function introduced in Eqn. (\ref{eq:loss_est}). 
\vspace{-0.1in}

\section{Approximating the IO by use of MCMC methods}
\label{ap:MCMC-IO}
In this section, a reference method based on MCMC techniques is described. This method can approximate the IO for a general signal detection-estimation. 
The difference between the hybrid method and the MCMC-based method is that the latter one employ MCMC techniques to approximate the likelihood ratio $\Lambda(\mathbf{g})$ and ideal estimate $\hat{\bm{\theta}}_I(\mathbf{g})$ while the former one use a multi-task CNN instead. The steps for approximating $\Lambda(\mathbf{g})$ and $\hat{\bm{\theta}}_I(\mathbf{g})$ are described below.
\vspace{-0.1in}

\subsection{Approximating the likelihood ratio}
Kupinski \etal $\>$proposed a method to approximate $\Lambda(\mathbf{g})$ numerically by employing MCMC techniques \cite{kupinski2003ideal} and Park \etal$\>$ extended the MCMC approach to signal-known-statistically (SKS) signal detection tasks \cite{park2003ideal}. Using the notations defined in Sec. \ref{sec:method}, the likelihood ratio $ \Lambda(\mathbf{g})= \frac{p(\mathbf{g}|H_1)}{p(\mathbf{g}|H_0)}$ can be written as:
\begin{equation}
\begin{aligned}
\Lambda(\mathbf{g})&=\frac{\int d\mathbf{b}(\bm{\alpha})\int d\mathbf{s}(\bm{\theta})p(\mathbf{b}(\bm{\alpha}))p(\mathbf{g|b(\bm{\alpha}),s(\bm{\theta})},H_1)p(\mathbf{s}(\bm{\theta}))}{\int d\mathbf{b^\prime(\bm{\alpha})} p({\mathbf{b^\prime(\bm{\alpha})}}) p(\mathbf{g|b^\prime(\bm{\alpha})},H_0)}\\
&\equiv \int d\bm{\theta}\int d\bm{\alpha}\Lambda_{BSKE}(\mathbf{g|b(\bm{\alpha}),s(\bm{\theta})})p(\mathbf{\bm{\alpha}|g},H_0)p(\mathbf{\bm{\theta}}),
\end{aligned}
\end{equation}
where $\Lambda_{BSKE}(\mathbf{g|b,s})=\frac{p(\mathbf{g|b,s},H_1)}{p(\mathbf{g|b},H_0)}$ is the background and signal-known exactly (BSKE) likelihood ratio and $p(\mathbf{b|g},H_0)=\frac{p(\mathbf{g|b},H_0)p(\mathbf{b})}{\int d\mathbf{b^\prime} p(\mathbf{g|b^\prime},H_0)p(\mathbf{b^\prime})}$. The likelihood ratio can be subsequently approximated as:
\begin{equation}
\label{eq:Lambda}
\Lambda(\mathbf{g})=\frac{1}{J}\sum_{j=1}^{J} \Lambda_{BSKE}(\mathbf{g|b(\bm{\alpha^{(j)}}),s(\bm{\theta^{(j)}})}),
\end{equation}
where $(\bm{\theta}^{(j)},\bm{\alpha}^{(j)})$ are sampled from the distribution $p(\bm{\alpha}|\mathbf{g},H_0)p(\bm{
\theta})$. A Markov chain with initial parameter vectors $(\bm{\theta}^{(0)},\bm{\alpha}^{(0)})$ and proposal densities $q_1(\bm{\theta}|\bm{\theta}^{(j)})$ and $q_2(\bm{\alpha}|\bm{\alpha}^{(j)})$ is constructed to generate $(\bm{\theta}^{(j)},\bm{\alpha}^{(j)})$, which is described below.
Given $(\bm{\theta}^{(i)},\bm{\alpha}^{(i)})$, candidate parameter vectors $(\bm{\theta}^\ast,\bm{\alpha}^\ast)$ are drawn from the proposal density and are accepted to the Markov chain with the acceptance probability:
\begin{equation}
\begin{aligned}
\label{eq:accp_lambda}
&p_a(\bm{\theta}^\ast,\bm{\alpha}^\ast|\bm{\theta^{(i)}},\bm{\alpha}^{(i)},\mathbf{g})=\\
&\min \left[1, \frac{p(\mathbf{g}|\mathbf{b}(\bm{\alpha}^\ast),H_0)p(\bm{\theta}^\ast)p(\bm{\alpha}^\ast)q_1(\bm{\theta}^\ast|\bm{\theta}^{(i)})q_2(\bm{\alpha}^\ast|\bm{\alpha}^{(i)})}{p(\mathbf{g}|\mathbf{b}(\bm{\alpha}^{(i)}),H_0)p(\bm{\theta}^{(i)})p(\bm{\alpha}^{(i)})q_1(\bm{\theta}^{(i)}|\bm{\theta}^\ast)q_2(\bm{\alpha}^{(i)}|\bm{\alpha}^\ast)} \right].
\end{aligned}
\end{equation}
The vectors $(\bm{\theta}^{(i+1)},\bm{\alpha}^{(i+1)})\equiv(\bm{\theta}^\ast,\bm{\alpha}^\ast)$ if the candidates are accepted; otherwise $(\bm{\theta}^{(i+1)},\bm{\alpha}^{(i+1)})\equiv(\bm{\theta}^{(i)},\bm{\alpha}^{(i)})$. When a random walk Metropolis-Hastings (RWMH) \cite{pereyra2015survey} is used here, the proposal density $q(\bm{\theta}|\bm{\theta}^{(j)})$ can be a simple Gaussian density. Therefore, $\Lambda(\mathbf{g})$ can be approximated.
\vspace{-0.15in}

\subsection{Approximating the ideal estimate}
Here, the utility function is assumed to be a quadratic utility function $u(\hat{\bm{\theta}},\bm{\theta})=1-\frac{1}{\epsilon}\Vert\hat{\bm{\theta}}-\bm{\theta}\Vert_2^2$. According to Eqn. (\ref{eq:IO_test_estimate_est}), the ideal estimator can be written as:
\begin{equation}
\hat{\bm{\theta}}_I(\mathbf{g})=\argmax_{\hat{\bm{\theta}}} \left\{\int p(\bm{\theta})\frac{p(\textbf{g}|\bm{\theta},H_1)}{p(\textbf{g}|H_0)}u(\bm{\theta},\hat{\bm{\theta}})d\bm{\theta}\right\}.
\end{equation}
when background variability is considered, given that the density of background is independent to the hypothesis, i.e., $p(\mathbf{b})=p(\mathbf{b}|H_1)$ and $p(\bm{\theta})=p(\bm{\theta}|H_1)$, the optimization can be written as following by using the Bayes' rule after term cancellations:
\begin{equation}
\hat{\bm{\theta}}_I(\mathbf{g})=\argmax_{\hat{\bm{\theta}}} \left\{\int d\bm{\theta} \int d\bm{\alpha} p(\bm{\theta},\mathbf{b}(\bm{\alpha})|\mathbf{g},H_1)u(\bm{\theta},\hat{\bm{\theta}})\right\}.
\end{equation}
Considering a quadratic based utility function:
\begin{equation}
\label{eq:optimization_appendix0}
\hat{\bm{\theta}}_I(\mathbf{g})=\argmin_{\hat{\bm{\theta}}} \left\{\int d\bm{\theta} \int d\bm{\alpha} p(\bm{\theta},\mathbf{b}(\bm{\alpha})|\mathbf{g},H_1)\Vert\hat{\bm{\theta}}-\bm{\theta}\Vert_2^2\right\}.
\end{equation}
By taking the derivative of Eqn. (\ref{eq:optimization_appendix0}) with respect to ${\hat{\bm{\theta}}}$ and setting the derivative to zero, $\hat{\bm{\theta}}_I(\mathbf{g})$ can be expressed as:
\begin{equation}
\label{eq:optimization_appendix}
\hat{\bm{\theta}}_I(\mathbf{g})=\iint \bm{\theta}p(\bm{\theta},\mathbf{b}(\bm{\alpha})|\mathbf{g},H_1)d\bm{\theta}d\bm{\alpha}.
\end{equation}
$\hat{\bm{\theta}}_I(\mathbf{g})$ can be approximated by using Monte Carlo integration subsequently:
\begin{equation}
\label{eq:Ug}
\hat{\bm{\theta}}_I(\mathbf{g}) =\frac{1}{J}\sum_{j=1}^{J} \bm{\theta}^{(j)}.
\end{equation}
Here, $(\bm{\theta}^{(j)},\bm{\alpha}^{(j)})$ are sampled from the posterior distribution $p(\bm{\theta},\mathbf{b}(\bm{\alpha})|\mathbf{g},H_1)$. 
A Markov chain with the initial parameter vectors $(\bm{\theta}^{(0)},\bm{\alpha}^{(0)})$ and the proposal densities $q_1(\bm{\theta}|\bm{\theta}^{(j)})$ and $q_2(\bm{\alpha}|\bm{\alpha}^{(j)})$ is constructed to generate $(\bm{\theta}^{(j)},\bm{\alpha}^{(j)})$, which is described below. 
Given $(\bm{\theta}^{(i)},\bm{\alpha}^{(i)})$, candidate parameter vectors $(\bm{\theta}^\ast,\bm{\alpha}^\ast)$ are drawn from the proposal density and are accepted to the Markov chain with the acceptance probability:
\begin{equation}
\begin{aligned}
&p_a(\bm{\theta}^\ast,\mathbf{b}(\bm{\alpha}^\ast)|\bm{\theta^{(i)}},\mathbf{b}(\bm{\alpha}^{(i)}),\mathbf{g})=\\
&\min \left[ 1,\frac{p(\mathbf{g}|\bm{\theta}^\ast,\mathbf{b}(\bm{\alpha}^\ast),H_1)p(\bm{\theta}^\ast)p(\bm{\alpha}^\ast)q_1(\bm{\theta}^\ast|\bm{\theta}^{(i)})q_2(\bm{\alpha}^\ast|\bm{\alpha}^{(i)})}{p(\mathbf{g}|\bm{\theta}^{(i)},\mathbf{b}(\bm{\alpha}^{(i)}),H_1)p(\bm{\theta}^{(i)})p(\bm{\alpha}^{(i)})q_1(\bm{\theta}^{(i)}|\bm{\theta}^\ast)q_2(\bm{\alpha}^{(i)}|\bm{\alpha}^\ast)}\right].
\end{aligned}
\end{equation}
The vectors $(\bm{\theta}^{(i+1)},\bm{\alpha}^{(i+1)})\equiv(\bm{\theta}^\ast,\bm{\alpha}^\ast)$ if the candidates are accepted; otherwise $(\bm{\theta}^{(i+1)},\bm{\alpha}^{(i+1)})\equiv(\bm{\theta}^{(i)},\bm{\alpha}^{(i)})$. When a random walk Metropolis-Hastings (RWMH) \cite{pereyra2015survey} is used here, the proposal density can be a simple Gaussian density. Therefore, $\hat{\bm{\theta}}_I(\mathbf{g})$ can be approximated.
\vspace{-0.0611in}
\bibliography{EROC} 
\bibliographystyle{IEEETran}

\end{document}